\documentclass[9pt,conference]{IEEEtran}
\IEEEoverridecommandlockouts

\usepackage{cite}
\usepackage{amsmath,amssymb,amsfonts}
\usepackage{algorithmic}
\usepackage{graphicx}
\usepackage{textcomp}
\usepackage{xcolor}
\usepackage{svg}
\usepackage{tipa}
\usepackage{colortbl}
\usepackage{multirow,arydshln}
\usepackage{tablefootnote}
\usepackage{balance}
\usepackage{hyperref}
\usepackage[linesnumbered, ruled, vlined]{algorithm2e}

\usepackage{xcolor,soul}

\def\BibTeX{{\rm B\kern-.05em{\sc i\kern-.025em b}\kern-.08em
    T\kern-.1667em\lower.7ex\hbox{E}\kern-.125emX}}
\begin{document}

\title{Semi-Supervised Cognitive State Classification from Speech with Multi-View Pseudo-Labeling}

\makeatletter
\newcommand{\linebreakand}{%
  \end{@IEEEauthorhalign}
  \hfill\mbox{}\par
  \mbox{}\hfill\begin{@IEEEauthorhalign}
}
\makeatother


\author{
\IEEEauthorblockN{Yuanchao Li$^1$, Zixing Zhang$^2$, Jing Han$^3$, Peter Bell$^1$, Catherine Lai$^1$}
\IEEEauthorblockA{\textit{$^1$University of Edinburgh, UK, $^2$Hunan University, China, $^3$University of Cambridge, UK}\\yuanchao.li@ed.ac.uk}
}

\maketitle

\begin{abstract}
The lack of labeled data is a common challenge in speech classification tasks, particularly those requiring extensive subjective assessment, such as cognitive state classification. In this work, we propose a Semi-Supervised Learning (SSL) framework, introducing a novel multi-view pseudo-labeling method that leverages both acoustic and linguistic characteristics to select the most confident data for training the classification model. Acoustically, unlabeled data are compared to labeled data using the Fréchet audio distance, calculated from embeddings generated by multiple audio encoders. Linguistically, large language models are prompted to revise automatic speech recognition transcriptions and predict labels based on our proposed task-specific knowledge. High-confidence data are identified when pseudo-labels from both sources align, while mismatches are treated as low-confidence data. A bimodal classifier is then trained to iteratively label the low-confidence data until a predefined criterion is met. We evaluate our SSL framework on emotion recognition and dementia detection tasks. Experimental results demonstrate that our method achieves competitive performance compared to fully supervised learning using only 30\% of the labeled data and significantly outperforms two selected baselines.
\end{abstract}


\section{Introduction}
\label{sec:intro}
Speech classification tasks, especially those related to cognitive states, are crucial for various applications, including human-computer interaction, health monitoring, and clinical diagnosis. However, human annotation for these tasks is often expensive, time-consuming, and requires extensive subjective assessment. This data scarcity issue hinders the progress of these classifications in real-world applications. To address this challenge, Semi-Supervised Learning (SSL) offers a promising approach by leveraging both limited labeled data and a larger amount of unlabeled data for various classification tasks, such as hate speech detection, emotion recognition, sound event detection, sleepiness detection, and gender classification \cite{d2020label,zhang2012semi,zhang2013co}.

Generally, SSL methods can be categorized into two main types: generating \textit{reliable pseudo-labels} and building \textit{reliable models} using limited data. For example, pseudo-labeling involves creating labels for unlabeled data based on predictions from an iteratively trained model \cite{zhu2021speech}. Consistency regularization ensures that the model produces consistent predictions for augmented versions of the same input (e.g., with added noise) \cite{lu2019semi}. These two types are often complementary and are commonly used together in existing SSL frameworks \cite{feng2022semi}. Moreover, the advancements in self-supervised learning have further improved SSL by utilizing large amounts of unlabeled data to pre-train SSL models \cite{zhang2022censer,lai2021semi}.

Among the literature, the most relevant studies to our work focus on pseudo-label generation. D'Sa et al. \cite{d2020label} used label propagation, transducing labels from labeled data to unlabeled data with probabilistic transitions for hate speech classification. Zhang et al. \cite{zhang2012semi} added unlabeled data with high confidence levels to the training set and resampled the originally labeled data for sound event classification. They also proposed dividing acoustic features into two views, selecting high-confidence instances in each view, and aggregating them with their predictions into the initial training sets per iteration \cite{zhang2013co}. Zhu et al. \cite{zhu2021speech} applied noisy student training \cite{xie2020self} to emotion recognition, using a teacher model trained on labeled data to infer soft labels for unlabeled data. Feng et al. \cite{feng2022semi} proposed incorporating federated learning, utilizing both labeled and unlabeled data at each local client in a multi-view pseudo-labeling approach.

Despite these advances, selecting high-confidence pseudo-labeled data remains challenging. In this work:

$\bullet$ We propose a novel SSL framework, integrating multi-view pseudo-labeling that leverages both acoustic and linguistic characteristics to select the most confident data for model training.

$\bullet$ We employ Fréchet Audio Distance (FAD) as a reference-free method to cluster unlabeled data based on acoustic similarity.

$\bullet$ We use task-specific prompts to predict labels from ASR transcripts, learning insights from acoustics, linguistics, and psychology.

$\bullet$ We examine multiple fusion methods in the context of SSL to build the bimodal classifier.

\begin{figure*}[th]
    \centering
    \includegraphics[width=0.98\textwidth]{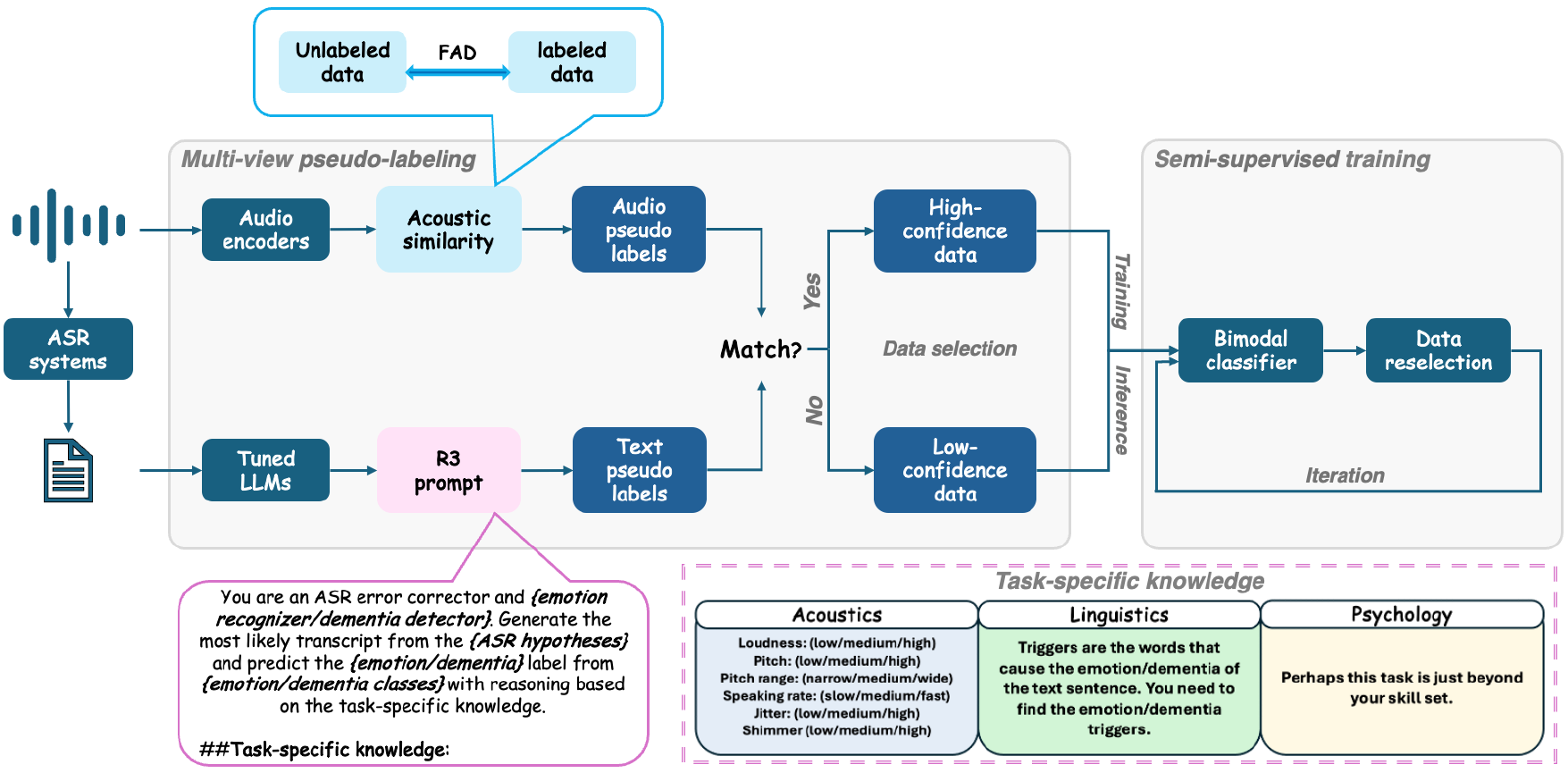}
    \caption{Framework of our proposed semi-supervised speech classification with pseudo-labeling and task-specific knowledge.}
    \label{fig:framework}
\end{figure*}

Our proposed SSL framework is evaluated on emotion recognition and dementia detection for classifying short-term and long-term cognitive states, demonstrating competitive performance using only 30\% of the labeled data compared to fully supervised learning, and showing greater effectiveness than the selected baselines.


\section{Methodology}

As illustrated in Fig.~\ref{fig:framework}, the proposed SSL framework with multi-view pseudo-labeling consists of two paths: acoustic and linguistic. The acoustic path utilizes the similarity between labeled and unlabeled data based on diverse audio embeddings, while the linguistic path employs LLMs to predict class labels from ASR transcriptions using task-specific knowledge. If the generated pseudo-labels from both paths align, we consider them as high-confidence data for training a bimodal classifier. Otherwise, the data are treated as low-confidence and will be further predicted using the trained bimodal classifier. The semi-supervised training of the bimodal classifier will iterate until a predefined criterion is met.

\subsection{Multi-View Pseudo-Labeling}
\subsubsection{Acoustic Path}
We extract acoustic features from multiple audio encoders trained with different objectives to reduce the bias of relying on a single one, inspired by a previous study in the music field \cite{li2024rethinking}. We use the following four audio encoders, resulting in four sets of embeddings for calculating the respective FAD scores:

$\bullet$ \textit{VGGish}: convolutional embeddings \cite{hershey2017cnn}

$\bullet$ \textit{EnCodec}: low-rate audio codecs \cite{defossezhigh}

$\bullet$ \textit{Wav2vec 2.0}: self-supervised acoustic embeddings \cite{baevski2020wav2vec}

$\bullet$ \textit{CLAP}: contrastive audio-text embeddings \cite{elizalde2024natural}

Given the embeddings of labeled data $\mathbf{X}^{l}$ and unlabeled data $\mathbf{X}^{u}$, the FAD score is calculated using multivariate Gaussians from two embedding sets $X^{l}(\mu_l, \Sigma_l)$ and $X^{u}(\mu_u, \Sigma_u)$ as follows:
\begin{align}
F(X^{l}, X^{u}) = ||\mu_l - \mu_u|| ^2 + tr(\Sigma_l + \Sigma_u - 2\sqrt{\Sigma_l\Sigma_u})
\end{align}
where $tr$ is the trace of a matrix.

Compared to traditional similarity metrics, such as cosine similarity and Euclidean distance, FAD is specifically designed for audio assessment, reflecting the perceptual similarity between two audio embedding distributions \cite{roblek2019fr}. It has proven effective in distinguishing real and synthetic audio \cite{gui2024adapting}, classifying audio with different emotions \cite{li2024rethinking}, and measuring acoustic similarity between emotional music and speech \cite{sun2024revisiting}. Therefore, we use FAD to measure the acoustic similarity between labeled and unlabeled data. The four FAD scores, calculated based on four sets of embeddings, are averaged to obtain the final FAD score. The unlabeled data with the smallest FAD score relative to the labeled class will be assigned that class label as the \textit{acoustic pseudo-labels}. An example of this process for emotion pseudo-labeling is shown in Table~\ref{tab:audio}. Note that the scores are not comparable across different audio encoders, as they are calculated based on different embeddings.

\begin{table}
\centering
\caption{An example of emotion pseudo-labeling using FAD. \textbf{Bold}: the smallest average FAD score, indicating its pseudo-label (i.e., Angry in this case).} 
\label{tab:audio}
\scalebox{1.05}{
\begin{tabular}{c|c|c|c|c}
\hline
 & \textbf{Angry} & \textbf{Happy} & \textbf{Neutral} & \textbf{Sad} \\ \hline
\textbf{VGGish} & 4.12 & 3.98 & 6.87 & 12.20 \\
\textbf{EnCodec} & 35.33 & 42.56 & 57.24 & 89.65 \\
\textbf{Wav2vec 2.0} & 54.66 & 58.49 & 88.78 & 109.02 \\
\textbf{CLAP} & 45.46 & 182.65 & 141.75 & 230.39 \\ \hdashline
\textbf{Average} & \textbf{34.64} & 71.42 & 73.41 & 110.57 \\ \hline
\end{tabular}}
\vspace{-5pt}
\end{table}

\subsubsection{Linguistic Path}
Previous speech classification tasks that used textual information typically relied on ground-truth text. However, in real-world applications, ASR is the only text source, and its transcriptions are usually noisy and contain errors, which can lead to incorrect classifications or pseudo-labels. We argue that \textit{it is more challenging to prompt LLMs for classification tasks based on ASR transcriptions compared to human transcriptions due to the presence of word errors.}

To address this, we use the \textsc{Revise-Reason-Recognize} (R3) prompting pipeline to perform speech classification with ASR Error Correction (AEC) and reasoning on ASR transcriptions \cite{li2024revise}. The R3 pipeline involves three steps: \textsc{Revise}, where ASR errors are corrected based on N-best hypotheses; \textsc{Reason}, where the LLMs self-explain based on the corrected transcriptions and task-specific knowledge; and \textsc{Recognize}, where the label is identified.

For the ASR systems, we adopt the following ten models, the same as \cite{li2024speech}, to generate diverse transcriptions and form 10-best ASR hypotheses:

$\bullet$ \textit{Wav2Vec2-base-\{100h,960h\}}

$\bullet$ \textit{Wav2Vec2-large-960h}

$\bullet$ \textit{Wav2Vec2-large-960h-lv60-self}

$\bullet$ \textit{HuBERT-large-ls960-ft}

$\bullet$ \textit{WavLM-libri-clean-100h-base-plus}

$\bullet$ \textit{Whisper-\{tiny, base, small, large-v2\}.en}

To perform AEC, we follow an AEC-specific Alpaca template \cite{yang2023generative}, which uses the \textit{``You are an ASR error corrector''} instruction, guiding the LLMs to perform error correction. As LLMs have demonstrated their ability in both AEC and emotion recognition \cite{yang2024large}, we expect that this capability can be extended to dementia detection from ASR transcriptions as well. The revised ASR transcriptions will be used for subsequent text feature extraction to train the bimodal classifier. For reasoning, we design task-specific knowledge that incorporates acoustics, linguistics, and psychology as in Fig.~\ref{fig:framework}.

We first apply Parameter-Efficient Fine-Tune (PEFT) on the following three LLMs with the LoRA adapter \cite{hulora} using the labeled data with the R3 prompt:

$\bullet$ \textit{Llama2-7b-chat-hf}

$\bullet$ \textit{Llama2-13b-chat-hf}

$\bullet$ \textit{Falcon-7b-instruct}

The learning rate, weight decay, and number of epochs are set to 1e-4, 1e-5, and 5, respectively, with AdamW optimizer used. The three fine-tuned LLMs are then prompted with the R3 prompt to predict class labels from ASR transcriptions of the unlabeled data. Finally, majority voting is applied to the predicted labels of the three LLMs to generate the \textit{linguistic pseudo-labels}.

\subsubsection{Data Selection}
The acoustic and linguistic pseudo-labels generated from the two paths are combined to select the most confident data for semi-supervised training. Data with matching pseudo-labels from both paths are selected as high-confidence, while data with differing pseudo-labels are considered low-confidence. Together with the labeled data, the high-confidence data will be used to train the bimodal classifier in the first iteration, ensuring robust initial training.

\subsection{Semi-Supervised Training}
\subsubsection{Bimodal Classifier}

The bimodal classifier consists of pre-trained feature encoders—\textit{HuBERT} \cite{hsu2021hubert} and \textit{RoBERTa} \cite{liu2019roberta}—that extract audio and text features, respectively, and a classification model that uses these features to generate a prediction label. For PEFT the encoders and training the classification model, the learning rate, weight decay, number of epochs, and batch size are set as 1e-4, 1e-5, 30, and 64, respectively. The AdamW optimizer is used. For the classification model, we examine the following four fusion methods:

\begin{itemize}
    \item \textit{Early fusion}: text and audio features are concatenated at the embedding level
    \item \textit{Cross-attention fusion}: text and audio features are attended to each other via attention and then concatenated \cite{li2022fusing}
    \item \textit{Tensor fusion}: unimodal information and bimodal interactions are learned explicitly and then aggregated \cite{zadeh2017tensor}
    \item \textit{Modality-gated fusion}: primary modality is dynamically adjusted in each training step \cite{wang2023cross}
\end{itemize}

\subsubsection{Iteration}
After training the bimodal classifier, low-confidence data are predicted and labeled using the trained classifier. In most previous SSL studies, model-labeled data are fully trusted and incorporated into the training set in the next iteration. However, as training progresses, mislabeled data (noise) may accumulate, leading to a cycle of erroneous learning \cite{zhu2005semi}. To address this issue, we choose not to fully trust the model-labeled data. Instead, the pseudo-label generated by the bimodal classifier is compared with pseudo-labels from multi-view pseudo-labeling. If the model pseudo-label matches either the acoustic or linguistic pseudo-label, the data are added to the training set for the next iteration. Otherwise, they remain low-confidence and will be predicted in the next iteration.

In each iteration, we update the training set by adding model-labeled data and randomly removing 20\% of the initial high-confidence data to avoid over-reliance on multi-view pseudo-labeling. The model is reinitialized in every iteration to reduce overfitting and bias. The maximum number of iterations is set to 40. However, if there is no performance improvement on the validation set for two consecutive iterations, the iteration will be terminated. The process is summarized in Algorithm~\ref{alg:iteration}.

\begin{algorithm}
    \small
    \caption{Iteration Process}
    \label{alg:iteration}
    \SetKwInOut{Input}{Input}
    \Input{
      $H$: Bimodal classifier; $S_v$: Validation set; $D_c$: Confident (high-confidence) data; $D_u$: Unconfident (low-confidence) data; $L^a$: Acoustic pseudo-labels; $L^l$: Linguistic pseudo-labels; $L^h$: Model pseudo-labels; $I$: Maximum number of iterations
    }
    \For{$i = 1, \dots, I$}
    {
        Train classifier $H^i \gets f(D^i_c)$\;
        Evaluate $L_e \gets H^i(S_v)$\;
        \If{performance on $D_e$ does not improve}{
            \textbf{break}\;
        }
        Generate pseudo-label $L^i_h \gets h^i(D_u)$\;
        \If{$L^i_h$ equals $L^i_a$ \textbf{or} $L^i_l$}{
            $L^i_u \gets L^i_h$\;
        }
        Update $D^{i}_c$ and $D^{i}_u$\;
        Reinitialize classifier $H^i$\;
    }
    \textbf{end}
\end{algorithm}

\section{Experiments}

\begin{table*}
    \centering
    \caption{Performance comparison with \textit{supervised\_full} and \textit{supervised\_limited} (results in UA\%). $N$: ground-truth labeling rate of training data. \textbf{BOLD}: best performance in each ground-truth labeling rate.}
    \label{tab:results}
    \scalebox{1.08}{
    \begin{tabular}{llccccccc}
        \hline
         & \multirow{2}{*}{\textbf{Fusion}} & \multirow{2}{*}{\textbf{S\_full}} & \multicolumn{2}{c}{\textbf{$N=30\%$}} & \multicolumn{2}{c}{\textbf{$N=25\%$}} & \multicolumn{2}{c}{\textbf{$N=20\%$}} \\
         &  &  & {\textbf{S\_limited}} & {\textbf{Ours}} & {\textbf{S\_limited}} & {\textbf{Ours}} & {\textbf{S\_limited}} & {\textbf{Ours}} \\ \hline
        \multirow{4}{*}{\textbf{\begin{tabular}[c]{@{}l@{}}Emotion\\ recognition\end{tabular}}} & \textit{Early} & 73.67 & 70.01 & \textbf{72.52} (+2.51) & 69.76 & \textbf{72.04} (+2.28) & 68.89 & \textbf{71.20} (+2.31) \\
         & \textit{C-attn} & 74.02 & 71.20 & \textbf{73.79} (+2.59) & 69.97 & \textbf{72.87} (+2.90) & 69.03 & \textbf{71.85} (+2.82) \\
         & \textit{Tensor} & 75.18 & 72.00 & \textbf{74.79} (+2.79) & 70.42 & \textbf{73.25} (+2.83) & 69.52 & \textbf{72.34} (+2.82) \\
         & \textit{M-gated} & 75.53 & 72.03 & \textbf{75.10} (+3.07) & 71.19 & \textbf{73.90} (+2.71) & 70.88 & \textbf{73.61} (+2.73) \\ \hdashline
        \multirow{4}{*}{\textbf{\begin{tabular}[c]{@{}l@{}}Dementia\\ detection\end{tabular}}} & \textit{Early} & 80.03 & 75.93 & \textbf{79.04} (+3.11) & 73.55 & \textbf{78.75} (+5.20) & 71.87 & \textbf{78.10} (+6.23) \\
         & \textit{C-attn} & 80.32 & 76.77 & \textbf{79.99} (+3.22) & 73.76 & \textbf{78.89} (+5.13) & 71.18 & \textbf{78.25} (+7.07) \\
         & \textit{Tensor} & 80.71 & 76.62 & \textbf{80.12} (+3.50) & 74.82 & \textbf{79.48} (+4.66) & 72.01 & \textbf{78.83} (+6.82) \\
         & \textit{M-gated} & 81.23 & 77.14 & \textbf{80.87} (+3.73) & 74.78 & \textbf{79.71} (+4.93) & 72.14 & \textbf{79.11} (+6.97) \\ \hline
        \end{tabular}}
    \end{table*}

\subsection{Datasets and Experimental Settings}
We use IEMOCAP \cite{busso2008iemocap} for emotion recognition and ADReSSo \cite{luz2021detecting} for dementia detection. For IEMOCAP, we focus on the Big Four emotion classes and exclude utterances with blank transcriptions, resulting in 5,500 utterances (\textit{1,103 angry, 1,615 happy+excited, 1,704 neutral, 1,078 sad}). For ADReSSo, since there are no human labels in the test set to verify our approach, we use only the training set and focus on binary classes: Alzheimer's Dementia (AD) and Cognitively Normal (CN). Due to the long duration of each audio file and the presence of interviewer's speech, we segment all files using the official segmentation information, extracting participants' speech, which results in 2,268 utterances (\textit{1,200 AD, 1,068 CN}). Punctuation and extra whitespace are removed, and all text is converted to lowercase. The revised ASR transcriptions by R3 prompt yield word error rates of 11.48\% and 30.25\% for IEMOCAP and ADReSSo, respectively (the groundtruth transcription of ADReSSo are created by Whisper transcribing, followed by a native speaker's correction, as there is no ground truth provided in the dataset.).

For both tasks, we use an 80/10/10 split for training, validation, and testing, applying the ratio equally to each class to ensure a balanced distribution. Additionally, the ground-truth labeling rates for the training data are compared at 20\%, 25\%, and 30\%, with the remaining data labeled using our method. All results are measured using Unweighted Accuracy (UA). Random seeds are kept consistent across all experiments. Other settings have been detailed in the previous section (\href{https://github.com/yc-li20/Semi-supervised-training}{Code available}).

\subsection{Results and Discussions}
Four baselines are used for comparison:

\begin{itemize}
    \item \textit{Supervised\_full}: the classification model is trained on the entire training data (i.e., the 80\% split)
    \item \textit{Supervised\_limited}: the classification model is trained on the limited labeled data without pseudo-labeling the unlabeled data
    \item \textit{Decision merging}: two classification models are trained using audio and text, respectively, and their probability distribution are merged to select high-confidence data for the next iteration
    \item \textit{Co-training}: two classification models are trained using audio and text, respectively. High-confidence data selected by each model are added to the training set for the other model in the next iteration \cite{blum1998combining,zhang2018leveraging}
\end{itemize}

The comparison of our method with \textit{supervised\_full} and \textit{supervised\_limited} are shown in Table~\ref{tab:results}. It can be observed that \textbf{\textit{1)}} With only 30\% labeled data, our method achieves performance competitive with the \textit{supervised\_full} baseline. \textbf{\textit{2)}} Our proposed method outperforms the \textit{supervised\_limited} baseline, which lacks multi-view pseudo-labeling and semi-supervised training to augment the labeled data, particularly in dementia detection. \textbf{\textit{3)}} The less ground-truth labeled data available, the more effective our method is in dementia detection, likely because binary classes are easier for pseudo-labeling. \textbf{\textit{4)}} When ground-truth labels are limited, classification performance of the \textit{supervised\_limited} baseline for both tasks drops significantly compared to the \textit{supervised\_full} training. Our proposed method, however, mitigates this drop by more than 2\% in emotion recognition and 3\%-7\% in dementia detection. \textbf{\textit{5)}} Modality-gated fusion performs best among the four fusion methods. This is reasonable as it dynamically selects the primary modality contributing most to the classification tasks, thereby reducing the impact of ASR errors in the text modality.

As bimodal fusion does not apply to the \textit{decision merging} and \textit{co-training} baseline settings, we select our best-performing results for comparison. Note that the principle of \textit{decision merging} is the same as that of a fusion method—late fusion (or decision-level fusion) \cite{snoek2005early}. The difference is that we further use the merged probability to determine the high-confidence data for iteration, whereas late fusion directly outputs results based on the highest probability. Here, we refer to it as \textit{decision merging} to avoid confusion with late fusion, as we have used fusion techniques in the previous experiment. For both \textit{decision merging} and \textit{co-training}, we set the threshold at 0.5 for emotion recognition and 0.7 for dementia detection. For example, the fourth emotion class will be selected as the pseudo-label for an unlabeled data with a probability distribution of \{0.1, 0.1, 0.3, 0.5\}, and the unlabeled data will be added to the training set as high-confidence data.

\begin{table}
\centering
    \caption{Performance comparison with \textit{decision merging} and \textit{co-training} (results in UA\%). $N$: ground-truth labeling rate of training data. \textbf{BOLD}: best performance in each rate.}
    \label{tab:result2}
    \scalebox{1.08}{
\begin{tabular}{llcc}
\hline
 &  & \textbf{\begin{tabular}[c]{@{}c@{}}Emotion\\ recognition\end{tabular}} & \textbf{\begin{tabular}[c]{@{}c@{}}Dementia\\ detection\end{tabular}} \\ \hline
\multirow{3}{*}{\textbf{$N=30\%$}} & \textit{Decision merging} & 67.60 & 74.55 \\
 & \textit{Co-training} & 69.99 & 74.73 \\
 & \textit{Ours} & \textbf{75.10} & \textbf{80.87} \\ \hdashline
\multirow{3}{*}{\textbf{$N=25\%$}} & \textit{Decision merging} & 66.34 & 72.70 \\
 & \textit{Co-training} & 69.04 & 72.87 \\
 & \textit{Ours} & \textbf{73.90} & \textbf{79.71} \\ \hdashline
\multirow{3}{*}{\textbf{$N=20\%$}} & \textit{Decision merging} & 66.12 & 71.07 \\
 & \textit{Co-training} & 69.00 & 72.10 \\
 & \textit{Ours} & \textbf{73.61} & \textbf{79.11} \\ \hline
\end{tabular}}
\end{table}

From Table~\ref{tab:result2}, we can observe that: \textbf{\textit{1)}} Our method significantly outperforms the two baselines, likely for two reasons: first, it generates more high-confidence unlabeled data with pseudo-labels, which iteratively trains the classifier for performance improvement; second, our bimodal classifier takes both modalities as input during training. In contrast, although the two baselines also consider two modalities, their classifiers are separate, each relying on a single modality and ignoring the interrelatedness between them. \textbf{\textit{2)}} Although both \textit{decision merging} and \textit{co-training} consist of two classifiers that output respective labels, the latter performs better than the former, especially in emotion recognition. This result is plausible since \textit{decision merging} could potentially weaken the better prediction if the other probability is significantly incorrect. On the contrary, by incorporating high-confidence data judged by the other view into training set, \textit{co-training} enables the two models to learn from each other indirectly. Its relatively lower effectiveness in dementia detection is likely due to the low-quality audio, which makes one of the models less powerful.

\subsection{Effect of Multi-View Pseudo-Labeling Components}
We explore the contributions of each audio encoder and LLM in multi-view pseudo-labeling by keeping either the acoustic or linguistic path unchanged (i.e., full path) while adding an individual encoder from the other path. For brevity, Table~\ref{tab:ablation} presents the results of early fusion with 30\% labeled data omitting the other conditions. The results show that \textbf{\textit{1)}} The acoustic path contributes more than the linguistic path. \textbf{\textit{2)}} \textit{CLAP} and \textit{Falcon} perform best among the acoustic and linguistic encoders, respectively.

\begin{table}
    \centering
    \caption{Contribution of each component of the multi-view pseudo-labeling (results in UA\%).}
    \scalebox{1.08}{
    \begin{tabular}{lcc}
        \hline
         & \textbf{Emotion recognition} & \textbf{Dementia detection} \\ \hline
        \textbf{Two paths} & 72.52 & 79.04 \\ \hdashline
        \multicolumn{3}{l}{\textbf{Full linguistic path}} \\
        \textit{  + VGGish} & 68.21 & 72.04 \\
        \textit{  + EnCodec} & 65.75 & 70.76 \\
        \textit{  + Wav2vec 2.0} & 68.87 & 72.00 \\
        \textit{  + CLAP} & 69.41 & 72.62 \\ \hdashline
        \multicolumn{3}{l}{\textbf{Full acoustic path}} \\
        \textit{  + Llama2-7b} & 70.53 & 76.40 \\
        \textit{  + Llama2-13b} & 71.31 & 77.91 \\
        \textit{  + Falcon} & 72.00 & 78.88 \\ \hline
        \end{tabular}}
    \label{tab:ablation}
    \end{table}

\section{Conclusion}
In this work, we propose a novel semi-supervised learning framework that introduces a multi-view pseudo-labeling method leveraging both acoustic and linguistic characteristics. This method utilizes Fréchet audio distance and large language models to select the most reliable unlabeled data for augmenting the training set. Multiple fusion techniques have been compared to utilize multi-view knowledge for further enhancement of the framework. We evaluate our method on emotion recognition and dementia detection tasks, demonstrating that it outperforms fully-supervised, limited-supervised, and two SSL baselines. Our method achieves competitive performance compared to fully supervised learning while using less than 30\% of human-labeled data. In future work, we plan to explore the effects of additional audio encoders and large language models on multi-view pseudo-labeling and investigate more efficient fusion methods for the bimodal classifier.

\section*{Acknowledgment}
We thank Prof. Jiahong Yuan (USTC) for his feedback on the data processing of ADReSSo dataset.

\balance
\bibliographystyle{IEEEbib}
\bibliography{refs}

\begin{thebibliography}{10}

\bibitem{d2020label}
Ashwin~Geet d'Sa, Irina Illina, Dominique Fohr, Dietrich Klakow, and Dana Ruiter,
\newblock ``Label propagation-based semi-supervised learning for hate speech classification,''
\newblock in {\em Insights from Negative Results Workshop, EMNLP 2020}, 2020.

\bibitem{zhang2012semi}
Zixing Zhang and Bj{\"o}rn Schuller,
\newblock ``Semi-supervised learning helps in sound event classification,''
\newblock in {\em 2012 IEEE International Conference on Acoustics, Speech and Signal Processing (ICASSP)}. IEEE, 2012, pp. 333--336.

\bibitem{zhang2013co}
Zixing Zhang, Jun Deng, and Bj{\"o}rn Schuller,
\newblock ``Co-training succeeds in computational paralinguistics,''
\newblock in {\em 2013 IEEE International Conference on Acoustics, Speech and Signal Processing (ICASSP)}. IEEE, 2013, pp. 8505--8509.

\bibitem{zhu2021speech}
Zhi Zhu and Yoshinao Sato,
\newblock ``Speech emotion recognition using semi-supervised learning with efficient labeling strategies,''
\newblock in {\em 2021 IEEE Automatic Speech Recognition and Understanding Workshop (ASRU)}. IEEE, 2021, pp. 358--365.

\bibitem{lu2019semi}
Kangkang Lu, Chuan-Sheng Foo, Kah~Kuan Teh, Huy~Dat Tran, and Vijay~Ramaseshan Chandrasekhar,
\newblock ``Semi-supervised audio classification with consistency-based regularization.,''
\newblock in {\em INTERSPEECH}, 2019, vol.~1, pp. 3654--3658.

\bibitem{feng2022semi}
Tiantian Feng and Shrikanth Narayanan,
\newblock ``Semi-fedser: Semi-supervised learning for speech emotion recognition on federated learning using multiview pseudo-labeling.,''
\newblock in {\em INTERSPEECH}, 2022.

\bibitem{zhang2022censer}
Bowen Zhang, Songjun Cao, Xiaoming Zhang, Yike Zhang, Long Ma, and Takahiro Shinozaki,
\newblock ``Censer: Curriculum semi-supervised learning for speech recognition based on self-supervised pre-training,''
\newblock in {\em INTERSPEECH}, 2022.

\bibitem{lai2021semi}
Cheng-I Lai, Yung-Sung Chuang, Hung-Yi Lee, Shang-Wen Li, and James Glass,
\newblock ``Semi-supervised spoken language understanding via self-supervised speech and language model pretraining,''
\newblock in {\em 2021 IEEE International Conference on Acoustics, Speech and Signal Processing (ICASSP)}. IEEE, 2021, pp. 7468--7472.

\bibitem{xie2020self}
Qizhe Xie, Minh-Thang Luong, Eduard Hovy, and Quoc~V Le,
\newblock ``Self-training with noisy student improves imagenet classification,''
\newblock in {\em Proceedings of the IEEE/CVF conference on computer vision and pattern recognition}, 2020, pp. 10687--10698.

\bibitem{li2024rethinking}
Yuanchao Li, Azalea Gui, Dimitra Emmanouilidou, and Hannes Gamper,
\newblock ``Rethinking emotion bias in music via frechet audio distance,''
\newblock {\em arXiv preprint arXiv:2409.15545}, 2024.

\bibitem{hershey2017cnn}
Shawn Hershey, Sourish Chaudhuri, Daniel~PW Ellis, Jort~F Gemmeke, Aren Jansen, R~Channing Moore, Manoj Plakal, Devin Platt, Rif~A Saurous, Bryan Seybold, et~al.,
\newblock ``{CNN} architectures for large-scale audio classification,''
\newblock in {\em 2017 IEEE International Conference on Acoustics, Speech and Signal Processing (ICASSP)}. IEEE, 2017, pp. 131--135.

\bibitem{defossezhigh}
Alexandre D{\'e}fossez, Jade Copet, Gabriel Synnaeve, and Yossi Adi,
\newblock ``High fidelity neural audio compression,''
\newblock {\em Transactions on Machine Learning Research}, 2022.

\bibitem{baevski2020wav2vec}
Alexei Baevski, Yuhao Zhou, Abdelrahman Mohamed, and Michael Auli,
\newblock ``wav2vec 2.0: A framework for self-supervised learning of speech representations,''
\newblock {\em Advances in neural information processing systems}, vol. 33, pp. 12449--12460, 2020.

\bibitem{elizalde2024natural}
Benjamin Elizalde, Soham Deshmukh, and Huaming Wang,
\newblock ``Natural language supervision for general-purpose audio representations,''
\newblock in {\em 2024 IEEE International Conference on Acoustics, Speech and Signal Processing (ICASSP)}. IEEE, 2024, pp. 336--340.

\bibitem{roblek2019fr}
Dominik Roblek, Kevin Kilgour, Matt Sharifi, and Mauricio Zuluaga,
\newblock ``Fr{\'e}chet audio distance: A reference-free metric for evaluating music enhancement algorithms,''
\newblock in {\em INTERSPEECH}, 2019, pp. 2350--2354.

\bibitem{gui2024adapting}
Azalea Gui, Hannes Gamper, Sebastian Braun, and Dimitra Emmanouilidou,
\newblock ``Adapting frechet audio distance for generative music evaluation,''
\newblock in {\em 2024 IEEE International Conference on Acoustics, Speech and Signal Processing (ICASSP)}. IEEE, 2024, pp. 1331--1335.

\bibitem{sun2024revisiting}
Yujia Sun, Zeyu Zhao, Korin Richmond, and Yuanchao Li,
\newblock ``Revisiting acoustic similarity in emotional speech and music via self-supervised representations,''
\newblock {\em arXiv preprint arXiv:2409.17899}, 2024.

\bibitem{li2024revise}
Yuanchao Li, Yuan Gong, Chao-Han~Huck Yang, Peter Bell, and Catherine Lai,
\newblock ``Revise, reason, and recognize: Llm-based emotion recognition via emotion-specific prompts and asr error correction,''
\newblock {\em arXiv preprint arXiv:2409.15551}, 2024.

\bibitem{li2024speech}
Yuanchao Li, Peter Bell, and Catherine Lai,
\newblock ``Speech emotion recognition with {ASR} transcripts: A comprehensive study on word error rate and fusion techniques,''
\newblock {\em 2024 IEEE Spoken Language Technology Workshop (SLT). IEEE}, 2024.

\bibitem{yang2023generative}
Chao-Han~Huck Yang, Yile Gu, Yi-Chieh Liu, Shalini Ghosh, Ivan Bulyko, and Andreas Stolcke,
\newblock ``Generative speech recognition error correction with large language models and task-activating prompting,''
\newblock in {\em 2023 IEEE Automatic Speech Recognition and Understanding Workshop (ASRU)}. IEEE, 2023, pp. 1--8.

\bibitem{yang2024large}
Chao-Han~Huck Yang, Taejin Park, Yuan Gong, Yuanchao Li, Zhehuai Chen, Yen-Ting Lin, Chen Chen, Yuchen Hu, Kunal Dhawan, Piotr {\.Z}elasko, et~al.,
\newblock ``Large language model based generative error correction: A challenge and baselines for speech recognition, speaker tagging, and emotion recognition,''
\newblock {\em arXiv preprint arXiv:2409.09785}, 2024.

\bibitem{hulora}
Edward~J Hu, Phillip Wallis, Zeyuan Allen-Zhu, Yuanzhi Li, Shean Wang, Lu~Wang, Weizhu Chen, et~al.,
\newblock ``{LoRA}: Low-rank adaptation of large language models,''
\newblock in {\em International Conference on Learning Representations}, 2021.

\bibitem{hsu2021hubert}
Wei-Ning Hsu, Benjamin Bolte, Yao-Hung~Hubert Tsai, Kushal Lakhotia, Ruslan Salakhutdinov, and Abdelrahman Mohamed,
\newblock ``Hubert: Self-supervised speech representation learning by masked prediction of hidden units,''
\newblock {\em IEEE/ACM transactions on audio, speech, and language processing}, vol. 29, pp. 3451--3460, 2021.

\bibitem{liu2019roberta}
Y~Liu,
\newblock ``{Roberta}: A robustly optimized bert pretraining approach,''
\newblock {\em arXiv preprint arXiv:1907.11692}, 2019.

\bibitem{li2022fusing}
Yuanchao Li, Peter Bell, and Catherine Lai,
\newblock ``Fusing {ASR} outputs in joint training for speech emotion recognition,''
\newblock in {\em 2022 IEEE International Conference on Acoustics, Speech and Signal Processing (ICASSP)}. IEEE, 2022, pp. 7362--7366.

\bibitem{zadeh2017tensor}
Amir Zadeh, Minghai Chen, Soujanya Poria, Erik Cambria, and Louis-Philippe Morency,
\newblock ``Tensor fusion network for multimodal sentiment analysis,''
\newblock in {\em Proceedings of the 2017 Conference on Empirical Methods in Natural Language Processing}, 2017, pp. 1103--1114.

\bibitem{wang2023cross}
Yaoting Wang, Yuanchao Li, Paul~Pu Liang, Louis-Philippe Morency, Peter Bell, and Catherine Lai,
\newblock ``Cross-attention is not enough: Incongruity-aware dynamic hierarchical fusion for multimodal affect recognition,''
\newblock {\em arXiv preprint arXiv:2305.13583}, 2023.

\bibitem{zhu2005semi}
Xiaojin~Jerry Zhu,
\newblock ``Semi-supervised learning literature survey,''
\newblock {\em University of Wisconsin-Madison Department of Computer Sciences}, 2005.

\bibitem{busso2008iemocap}
Carlos Busso, Murtaza Bulut, Chi-Chun Lee, Abe Kazemzadeh, Emily Mower, Samuel Kim, Jeannette~N Chang, Sungbok Lee, and Shrikanth~S Narayanan,
\newblock ``{IEMOCAP}: Interactive emotional dyadic motion capture database,''
\newblock {\em Language resources and evaluation}, vol. 42, pp. 335--359, 2008.

\bibitem{luz2021detecting}
Saturnino Luz, Fasih Haider, Sofia de~la Fuente, Davida Fromm, and Brian MacWhinney,
\newblock ``Detecting cognitive decline using speech only: The adresso challenge.,''
\newblock in {\em INTERSPEECH}, 2021.

\bibitem{blum1998combining}
Avrim Blum and Tom Mitchell,
\newblock ``Combining labeled and unlabeled data with co-training,''
\newblock in {\em Proceedings of the eleventh annual conference on Computational learning theory}, 1998, pp. 92--100.

\bibitem{zhang2018leveraging}
Zixing Zhang, Jing Han, Jun Deng, Xinzhou Xu, Fabien Ringeval, and Bj{\"o}rn Schuller,
\newblock ``Leveraging unlabeled data for emotion recognition with enhanced collaborative semi-supervised learning,''
\newblock {\em IEEE Access}, vol. 6, pp. 22196--22209, 2018.

\bibitem{snoek2005early}
Cees~GM Snoek, Marcel Worring, and Arnold~WM Smeulders,
\newblock ``Early versus late fusion in semantic video analysis,''
\newblock in {\em Proceedings of the 13th annual ACM international conference on Multimedia}, 2005, pp. 399--402.

\end{thebibliography}

\end{document}